\documentclass[prb,twocolumn,groupedaddress,showpacs]{revtex4}
\usepackage[dvips]{graphics,color}
\usepackage{amsmath,bm,amsfonts}
\DeclareGraphicsExtensions{.eps}

\def\bD{{\bm D}}\def\bH{{\bm H}}\def\bL{{\bm L}}\def\bP{{\bm P}}
\def\bT{{\bm T}}
\def\cA{{\cal A}}\def\cC{{\cal C}}\def\cZ{{\cal Z}}
\def\cR{{\cal R}}

\def\Fr{\mathrm{Fr}}
\def\tk{\tilde{k}}\def\tm{\tilde{m}}
\def\ket#1{\mid~{\!\!\! #1\!\!}~\rangle}
\def\bra#1{\langle~{\!\!\! #1\!\!}~\mid}
\def\braket#1#2{\langle~{\!\! #1\!\!}~\mid~{\!\! #2\!}~\rangle}

\def\newpic#1{}
\def\ri{\mathrm{i}}
\def\re{\mathrm{e}}
\def\tq{\tilde{q}}

\begin{document}

\bibliographystyle{revtex}
\title{{\bf CARBON NANOTUBES BAND ASSIGNATION, TOPOLOGY, BLOCH STATES AND SELECTION RULES}}
\author{T. Vukovi\'c}
\email{tanja37@afrodita.rcub.bg.ac.yu}\homepage{http://www.ff.bg.ac.yu/qmf/qsg_e.htm}
\author{I. Milo\v sevi\'c and M. Damnjanovi\'c}
\affiliation{Faculty of Physics, University of Belgrade, POB 368,
Beograd 11001, Yugoslavia}
\date{\today}
\begin{abstract}
Various properties of the energy band structures (electronic, phonon,
etc.), including systematic band degeneracy, sticking and extremes,
following from the full line group symmetry of the single-wall carbon
nanotubes are established. The complete set of quantum numbers
consists of quasi momenta (angular and linear or helical) and
parities with respect to the $z$-reversal symmetries and, for achiral
tubes, the vertical plane. The assignation of the electronic bands is
performed, and the generalized Bloch symmetry adapted eigen functions
are derived. The most important physical tensors are characterized by
the same set of quantum numbers. All this enables application of the
presented exhaustive selection rules. The results are discussed by
some examples, e.g. allowed interband transitions, conductivity,
Raman tensor, etc.
\end{abstract}
\pacs{71.20.Tx, 63.22.+m, 61.48.+c} \maketitle

\section{Introduction}
It is well known that the single-wall carbon nanotubes~\cite{IIJIMA}
(SWCT) besides the translational periodicity along the tube axis
($z$-axis, by convention) possess a screw axis and pure rotational
symmetries. Consequently, in calculations of the electronic energy
band structure the conserved quantum numbers of
linear~\cite{ELBANDS-I} $k$, or helical~\cite{ELBANDS-Y} $\tilde{k}$,
quasi momenta together with $z$-projection of the orbital angular
momentum (related to rotational symmetries) are used. On the
contrary, the parity quantum numbers following from the full line
group symmetry~\cite{YITR} including horizontal $U$-axis and, in the
zig-zag and armchair cases, vertical and horizontal mirror and glide
planes, have not been used in band assignation. It is important to
complete this task, since it yields many important exact properties
of the electronic band structures, some of them being quite
independent of the model considered. Let us mention only the band
degeneracies, systematic van Hove singularities and the precise
selection rules relevant for the processes in nanotubes. Further,
some general predictions on the topology of band sticking follow from
the interplay between linear and helical quantum numbers, and may be
a priori predicted and used for the completely assigned bands.

\begin{figure}[hbt]\centering
 \includegraphics[1mm,1mm][6.5cm,6.5cm]{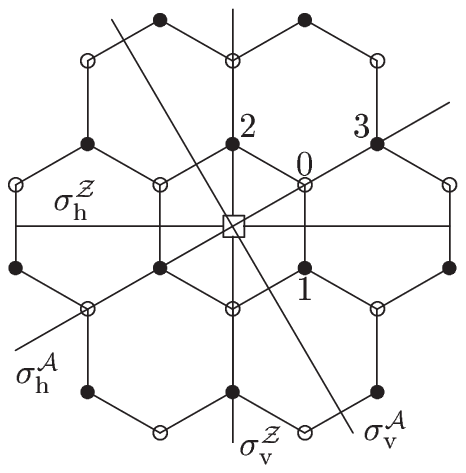}
 \caption{\label{Fswcneigh}{\bf Symmetry and neighbors.} Perpendicular
to the figure at center (box) is the $U$-axis (assumed to be the
$x$-axis), while $\sigma^{\cZ/\cA}_{\mathrm{h}/\mathrm{v}}$ stands
for vertical and horizontal mirror planes of $\cZ$ and $\cA$ tubes.
Atoms C$_{ts0}$ and C$_{ts1}$ are differed as $\circ$ and $\bullet$.
Nearest neighbors of the atom C$_{000}$, denoted by 0, are the atoms
1, 2 and 3.}
\end{figure}

All the geometrical symmetries of chiral $(n_1,n_2)$, zig-zag $(n,0)$
and armchair $(n,n)$ SWCT ($\cC$, $\cZ$ and $\cA$ tubes for short)
are gathered in the line groups~\cite{YITR} (the factorized and the
international notation are given):
\begin{subequations}\label{ELG}\begin{eqnarray}
 \bL_{\cC}&=&\bT^r_q\bD_n= \bL q_p22,\label{ELGC}\\
 \bL_{\cZ\cA}&=&\bT^1_{2n}\bD_{nh}=\bL 2n_n/mcm.\label{ELGZA}
\end{eqnarray}\end{subequations}
Here, $n$ is the greatest common divisor of $n_1$ and $n_2$,
$q=2(n^2_1+n_1n_2+n^2_2)/n{\cal R}$ with $\cR=3$ or $\cR=1$ whether
$(n_1-n_2)/3n$ is integer or not, while the helicity parameters $r$
and $p$ are expressed in terms of $n_1$ and $n_2$ by number
theoretical functions~\cite{QT}. The elements of the groups
\eqref{ELG} are (the coordinate system and the positions of the
symmetry axes and planes are presented in Fig. \ref{Fswcneigh}):
\begin{equation}\label{Elgel}
\ell(t,s,u,v)=(C^r_q|na/q)^tC^s_nU^u\sigma^v_x,
\end{equation}
where $(C^r_q|na/q)^t$ (Koster-Seitz notation; $a$ is the
translational period of the tube) for $t=0,\pm1,\dots$ are the
elements of the helical group (screw-axis) $\bT^r_q$. The rotations
$C^s_n$, $s=0,\dots,n-1$, around the $z$-axis form the subgroup ${\bm
C}_n$. Finally, $U$ is the rotation by $\pi$ around the $x$-axis
($u=0,1$), and $\sigma_{\mathrm{v}}$ the vertical mirror $xz$-plane
in the case of the achiral tubes, i.e. $v=0,1$ for $\cZ$ and $\cA$
tubes, and $v=0$ for $\cC$ ones. Each carbon atom on the tube is
obtained from a single one C$_{000}$ by the action of the element
$\ell(t,s,u,0)$. This enables to enumerate the atoms as C$_{tsu}$.
The isogonal point groups are:
\begin{equation}\label{Eisogroups}
\bP_{\cC}=\bD_q,\quad \bP_{\cZ\cA}=\bD_{2nh}.
\end{equation}

Each electronic eigen state corresponds to the complete set of
symmetry based quantum numbers, which singles out an irreducible
representation of the symmetry group of SWCT. Knowing this
representation the eigen states (in the form of the generalized Bloch
functions) and the eigen energies (organized as the energy bands) may
be easily found~\cite{ALTMAN,YTBA}. Analysis of the linear and
helical symmetry based quantum numbers and their mutual relations is
performed in section \ref{Sqn}; this enable to emphasize {\em a
priori} properties of the band structures, referring to any subsystem
(electrons, phonons, etc.). In section \ref{Sbands} these
considerations are further elaborated for the electronic bands: the
general dispersion relations are derived and especially in the most
usual tight-binding approach the complete symmetry assignation and
the Bloch eigen states are presented. Then, in section \ref{Sselrul},
possible applications in the analysis of different processes are
explained, using the general forms of various tensors (e.g.
dielectric permeability, Raman, conductivity)
and the selection rules
that are given in appendix \ref{Acgcoef}. Basic conclusions are
reviewed in the last section.

\section{Symmetry and quantum numbers}\label{Sqn}
For the chiral tubes, there are two sets of quantum numbers used in
literature: $km$- and $\tilde{k}\tilde{m}$-numbers. They correspond
to different parameterizations of the irreducible
representations~\cite{YTBA,IY-93} of the line groups ${\bf L}_{\cC}$.
The inherent physical meaning of the quantum numbers makes one or
another choice more suitable for different physical considerations.

The group structure \eqref{ELGC} immediately shows that the conserved
quantities~\cite{IY-93} are: the helical quasi momentum $\tilde{k}$
related to the screw-axis subgroup $\bT^r_q$, the $z$ component of
the angular quasi momentum $\tilde{m}$ arising from the pure
rotational symmetries ${\bm C}_n$ and the parity with respect to $U$
axis. Note that $\tk$ includes the linear quasi momentum $k$ and the
part of the angular momentum being not included in $\tm$. The
different values of $\tilde{k}$ are within the interval
$\widetilde{\mathrm{BZ}}=(-\tilde{\pi}/a,\tilde{\pi}/a]$ (here
$\tilde{\pi}=\tilde{q}\pi$; $\tilde{q}=q/n$), and $\tilde{m}$ runs
over the integers from $(-n/2,n/2]$. As usual, the equality of the
quantum numbers modulo these intervals is assumed. The parity quantum
number corresponding to the $U$-axis will be denoted as + and $-$ for
even and odd states. Thus, with $\tilde{k}\tilde{m}$-numbers the
states are labeled as $\ket{\tilde{k}\tilde{m}}$ or
$\ket{\tilde{k}\tilde{m}\pm}$ for the $U$-parity even and odd states.
Since the $U$-symmetry reverses the $z$-components both of the linear
and the angular momenta, the state $\ket{\tilde{k},\tilde{m}}$ is
mapped by $U$ to $\ket{-\tilde{k},-\tilde{m}}$, and vice versa. As
far as these two states are different, the corresponding energy
levels are degenerate, and the doublet spans a two dimensional space
carrying the irreducible representation ${_{\tilde{k}}}E_{\tilde{m}}$
with integer $\tilde{m}\in(-n/2,n/2]$. Therefore, it suffices to
consider $\tilde{k}$ only in the irreducible domain~\cite{ALTMAN}
$\widetilde{\mathrm{ID}}=[0,\tilde{\pi}/a]$. Obviously, only if
$k=0,\tilde{\pi}/a$ and $m=0,n/2$ the states
$\ket{\tilde{k},\tilde{m}}$ and $\ket{-\tilde{k},-\tilde{m}}$ are
physically the same:
$\ket{-\tilde{k},-\tilde{m}}=\pm\ket{\tilde{k}\tilde{m}}$. Thus, even
or odd states are at the edges~\cite{PI} of
$\widetilde{\mathrm{ID}}$: $\ket{00\pm}$, $\ket{\tilde{\pi}0\pm}$,
and only for even $n$ also $\ket{0,n/2,\pm}$ and
$\ket{\tilde{\pi},n/2,\pm}$. These nondegenerate states correspond to
the one-dimensional representations ${_0}A^\pm_0$,
${_{\tilde{\pi}}}A^\pm_0$, ${_0}A^\pm_{n/2}$ and
${_{\tilde{\pi}}}A^\pm_{n/2}$. So, the eigen energies make at least
double degenerate bands over the interior of
$\widetilde{\mathrm{ID}}$, and only the bands with $m=0,n/2$ end by
even or odd singlet state. As for the other bands, those with the
opposite $\tilde{m}$ are sticked together at
$\tilde{k}=0,\tilde{\pi}/a$.

With the $km$-numbers~\cite{YTBA}, a state of (quasi)particle
propagating along the $z$-axis with the quasi momentum $k$ and the
$z$-component of the angular momentum $m$ is denoted as $\ket{km}$,
or $\ket{km\pm}$. Reflecting pure translational symmetry, the
(linear) quasi momentum $k$ varies within the Brillouin zone
$\mathrm{BZ}=(-\pi/a,\pi/a]$, while $m$ is related to the isogonal
rotations and takes on the integer values from $(-q/2,q/2]$,
precisely\cite{QT} $m=-q/2+1,\dots,q/2$. It is assumed that the
equalities in $k$ and $m$ are modulo these intervals. Since the pure
rotation by $2\pi/q$ of the isogonal group (\ref{Eisogroups}) is not
a symmetry of the system, quantum number $m$ may not be conserved,
which is clearly manifested in the selection rules (see the
appendix). The states $\ket{km}$ and $\ket{-k,-m}$, being mapped one
onto another by the $U$-symmetry, make a degenerate doublet of the
irreducible representation ${_k}E_m$ for any $k$ in the interior of
the irreducible domain $\mathrm{ID}=[0,\pi/a]$. At the edge point
$k=0$, when $U$ leaves $k$ invariant, for $m=1,\dots,q/2-1$ the state
$\ket{0m}$ is mapped onto the state $\ket{0,-m}$, yielding again
two-dimensional irreducible representations ${_0}E_m$. Only for
$m=0,q/2$ there are nondegenerate even and odd states, $\ket{00\pm}$
and $\ket{0,q/2,\pm}$, carrying the one-dimensional irreducible
representations ${_0}A^\pm_0$ and ${_0}A^\pm_{q/2}$. At $k=\pi/a$,
$U$ intertwins the states $\ket{\pi m}$ and $\ket{\pi m'}$, where
$m'=-p-m$. Thus, the singlets~\cite{PI} $\ket{\pi,-p/2,\pm}$ and
$\ket{\pi,(q-p)/2,\pm}$ appear only for $p$ even~\cite{QT} and
correspond to the one-dimensional representations
${_\pi}A^\pm_{-p/2}$ and ${_\pi}A^\pm_{(q-p)/2}$. The remaining
integers $m\in(-p/2,(q-p)/2)$ give the double degenerate levels of
the representations ${_\pi}E_m$. Consequently, the bands with $m$ and
$m'$, differ in the interior of ID, but they stick together at
$k=\pi/a$ as well as those with opposite $m\neq 0,q/2$ are sticked
together at $k=0$.

The additional mirror planes $\sigma_{\mathrm{v}}$ and
$\sigma_{\mathrm{h}}=U\sigma_{\mathrm{v}}$ yield new parities in the
cases of $\cZ$ and $\cA$ tubes. Even and odd states with respect to
$\sigma_{\mathrm{v}}$ are labeled by $A$ and $B$. The parity of the
horizontal mirror plane $\sigma_{\mathrm{h}}$ is denoted as that of
$U$, i.e. '+' and '$-$' now points to the even and odd states with
respect to either one of these $z$-reversing operations. Obviously,
$\sigma_{\mathrm{v}}$ leaves $k$ invariant and $m$ is
reversed~\cite{HELICAL}. Therefore, in the interior of the ID, the
$U$-degenerate states $\ket{km}$ and $\ket{-k,-m}$ are mapped by
$\sigma_{\mathrm{v}}$ onto $\ket{k,-m}$ and $\ket{-k,m}$. For each
$m=1,\dots,n-1$ all these states span the four-dimensional
irreducible representation ${_k}G_m$ of the four fold degenerate
band. Only for $m=0,n$ the degeneracy remains two fold, in accordance
with the two dimensional irreducible representations ${_k}E^{A/B}_0$
and ${_k}E^{A/B}_n$ over $\sigma_{\mathrm{v}}$-even or odd states
$\ket{km,A/B}$ and $\ket{-km,A/B}$. If further $k=0$, the states
$\ket{00,\pm,A/B}$ and $\ket{0n,\pm,A/B}$ are nondegenerate,
corresponding to the one-dimensional representations ${_0}A^\pm_0$,
${_0}B^\pm_0$, ${_0}A^\pm_n$ and ${_0}B^\pm_n$. For the remaining
$m=1,\dots,n-1$, the states $\ket{0m\pm}$ and $\ket{0,-m\pm}$ are
degenerate giving two-dimensional representations ${_0}E^\pm_m$
(parity with respect to $\sigma_{\mathrm{h}}$). At the other ID edge
$k=\pi/a$, for integer $m\in(0,n/2)$ the four-fold degenerate states
$\ket{\pi m}$, $\ket{\pi,-m}$, $\ket{\pi,n-m}$ and $\ket{\pi,m-n}$
span the representation ${_\pi}G_m$. As for $m=0,n$, the states
$\ket{\pi 0,A/B}$ and $\ket{\pi n,A/B}$ as well as the states
(existing only for $n$ even) $\ket{\pi,n/2,\pm}$ and
$\ket{\pi,-n/2,\pm}$ are degenerate, being associated to the
representations ${_\pi}E^A_0$, ${_\pi}E^B_0$ and ${_\pi}E^\pm_{n/2}$.

\begin{figure}[hbt]\centering
  \includegraphics[1mm,1mm][8cm,9.7cm]{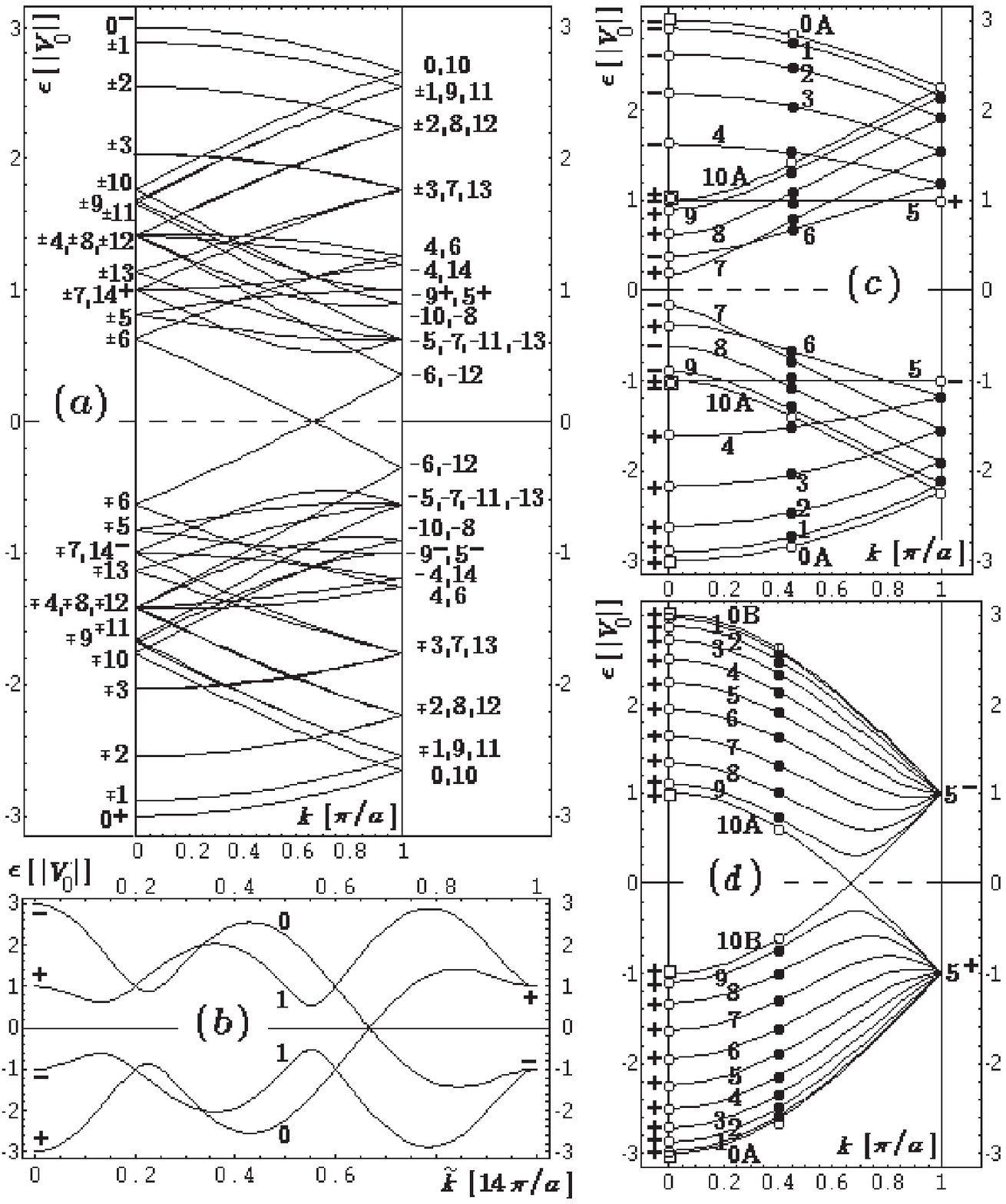}
  \caption[]{\label{Fcband}{\bf Symmetry assigned electronic bands of
SWCT}. For the chiral tube $(8,2)$ (line group $\bT^{11}_{28}{\bf
D}_2= \bL28_{18}22$, $a=\sqrt{7}a_0=6.5${\AA}) the bands are double
degenerate in the interior of ID, while at the edges the $U$-parity
 singlets are emphasized by $+$ or $-$; $km$- ($m$ is given at
both edges of the band ${_k}E_m$) and $\tilde{k}\tilde{m}$-numbers
assignation on $(a)$ and $(b)$. $(c)$ and $(d)$: The bands of the
zig-zag $(10,0)$ and the armchair $(10,10)$ tubes (line group
$\bT^{1}_{20}\bD_{10h}= {\bf L}20_{10}/mcm$,
$a_\cZ=\sqrt{3}a_0=4.26${\AA}, and $a_\cA=a_0$) are either four-fold
(${_k}G_m$, $\bullet$) or double degenerate (${_k}E^{A/B}_{0/10}$,
$\circ$, $\sigma_{\mathrm{v}}$ parity $A$ or $B$ given next to $m$);
$z$-reversal parity ($+$ or $-$) and nondegenerate states (box)
appear at the edges of ID.} \end{figure}

In terms of energies only, one concludes that the systematic band
degeneracy is caused by the parities only. Their nontrivial action on
the momenta eigen states defines the irreducible representations of
the symmetry group (\ref{ELG}) of the dimension 1 (representations
$A$ and $B$), 2 ($E$) or 4 ($G$), uniquely corresponding to each
complete set of all quantum numbers. The $U$-axis is manifested as
the symmetry of the bands with respect to $\tk=0,q\pi/na$:
\begin{subequations}\label{EesymU}\begin{eqnarray}\label{EesymU0}
 &&\epsilon_{\tm}(\tk)=\epsilon_{-\tm}(-\tk)=\epsilon_{-\tm}(2\tilde{\pi}/a-\tk),\\
 &&\epsilon_{\tm}(\tilde{\pi}/a-\tk)=\epsilon_{-\tm}(\tilde{\pi}/a+\tk).\label{EesymUpi}
\end{eqnarray}\end{subequations}
Two consequences, the possibility to consider only the irreducible
domain and the obligate double degeneracy in its interior have been
emphasized already. In addition, some general topological properties
of the bands can be predicted. The conditions (\ref{EesymU}) at
$\tk=0,\tilde{\pi}/a$ show that $\tm$ and $-\tm$ bands are sticked
together in the both $\widetilde{\mathrm{ID}}$ edges. These results
in the terms of $km$ numbers~\cite{HELICAL} become the bands $\pm m$
are sticked together at $k=0$, as well as the bands $m$ and $m+p$ at
$\pi/a$. Altogether, the irreducible domain $\widetilde{\mathrm{ID}}$
contains $q/n$ segments as wide as ID; the part of a
$\tilde{k}\tilde{m}$ band over each of the segments is one of the
$km$ bands having the quantum number $m=\tilde{m}+jn$
($j=0,\dots,\tilde{q}-1$). The successive segments correspond to the
sticked together $km$ bands. Further, the relations (\ref{EesymU})
imply van Hove singularities at $\tk=0,\tilde{\pi}/a$ of the
$\tilde{m}=0,n$ bands (when $\tilde{m}=-\tilde{m}$):
$[\mathrm{d}\epsilon_{\tilde{m}}(\tilde{k})/
\mathrm{d}\tilde{k}]_{\tilde{k}=0,\tilde{\pi}/a}=0$. The same
singularities~\cite{HELICAL} are found for $m=0,q/2$ at $k=0$ and
$m=-p/2,(q-p)/2$ at $k=\pi/a$. The importance of the coincidence of
these singularities with the $U$-parity even and odd states will be
discussed later on. Additional mirror planes of the achiral tubes
force $m$ and $-m$ bands to coincide, causing additional degeneracy,
together with the condition
\begin{equation}\label{EesymS}
 \epsilon_{\tilde{m}}(\tilde{k})=\epsilon_{\tilde{m}}(-\tilde{k})=
 \epsilon_{\tilde{m}}(2\tilde{\pi}/a-\tilde{k}).
\end{equation}
This yields the $z$-reversal parity of all the states at $k=0$ and
$m=n/2$ states at $k=\pi/a$, as well as in the corresponding van Hove
singularities. All these general conclusions are transparently
verified by the tight-binding electronic bands given in Fig.
\ref{Fcband}. For example, all the bands of the achiral tubes have
extremes at $k=0$ and only $m=5$ band also in $k=\pi/a$. Of course,
besides these systematic $z$-reversal parities caused extremes, other
ones may appear~\cite{REICH} depending on the considered model.

Let us remark that the same quantum numbers may characterize several,
say $N$ (frequency number~\cite{ALTMAN} of the corresponding
representation) eigen states with equal or different eigen energies.
In such cases the index $F$ differing between such states is added.
For example, for each $m$ there are $N=2$ ${_k}E_m$ electronic bands
of $\cC$-tubes (section \ref{Sbands}). These eigen energies
$\epsilon^\pm_m(k)$ and the vectors $\ket{km;\pm}$, $\ket{-k,-m;\pm}$
are distinguished by $F=\pm$.

\section{Electronic $\pi$-bands}\label{Sbands}
The tight binding hamiltonian including a single $\pi$-orbital
$\ket{tsu}$ per site C$_{tsu}$ is
\begin{equation*}
H=\sum_{tsu}\sum_{t's'u'}H_{tsu,t's'u'}\ket{tsu}\bra{t's'u'}.
\end{equation*}
The electronic band structure for such a hamiltonian has been
calculated (regarding nearest neighbor interactions) with the help of
the translational and the principle axis rotational symmetries. The
resulting assignation thus involved either $km$ or
$\tilde{k}\tilde{m}$-numbers~\cite{ELBANDS-I,ELBANDS-Y}. Here we
complete the assignation by the additional parities of the full
symmetry group. The results are obtained applying the modified group
projector technique~\cite{YTBA}.

Depending on the choice of the quantum numbers it is convenient to
introduce the phases:
\begin{eqnarray}\label{Ephase}
 &\psi^k_m(t,s)=\frac{kan+2\pi mr}{q}t+\frac{2\pi m}{n}s,\nonumber\\
 &\tilde{\psi}^{\tk}_{\tm}(t,s)=\frac{\tk a}{\tq}t+\frac{2\pi\tm}{n}s.
\end{eqnarray}
Then, for the generic bands of the chiral tube (double degenerate
${_k}E_m$ or ${_{\tilde{k}}}E_{\tilde{m}}$ bands for any allowed $m$
or $\tilde{m}$ along the interior of the irreducible domain)  the
dispersion relations and the corresponding eigen vectors are obtained
solving the eigen problem of
\begin{subequations}\label{EH|0C}\begin{eqnarray}\label{EH|0Ca}
 H_m(k)&=&\begin{pmatrix}h^0_m(k)&h^{1*}_m(k)\\
                       h^1_m(k)&h^0_m(k)\end{pmatrix},\\
 h^u_m(k)&=&\sum_{ts}H_{tsu}\re^{\ri\psi^m_k(t,s)} \quad (u=0,1),
\end{eqnarray}\end{subequations}
with $H_{tsu}=H_{000,tsu}$. Finally, for each $m$ one finds two bands
\begin{subequations}\label{EebooC}\begin{equation}\label{EeboovalC}
\epsilon^\pm_m(k)=h^0_m(k)\pm|h^1_m(k)|\end{equation}
with the corresponding generalized Bloch eigen functions
\begin{eqnarray}
 &&\ket{km;\pm}=\sum_{ts}\re^{-\ri\psi^{k}_{m}(t,s)}
  \big(\ket{ts0}\pm\re^{\ri h^{k}_{m}}\ket{ts1}\big),\\
 &&\ket{-k,-m;\pm}=\sum_{ts}\re^{\ri\psi^{k}_{m}(t,s)}
  \big(\ket{ts1}\pm\re^{\ri h^{k}_{m}}\ket{ts0}\big),
\label{EeboovecC}
\end{eqnarray}\end{subequations}
where $h^k_m=\mathrm{Arg}\big(h^1_m(k)\big)$. To get the expressions
in terms of $\tilde{k}\tilde{m}$ quantum numbers the angles
$\tilde{\psi}^{\tk}_{\tm}(t,s)$ are used to define
$\tilde{h}^u_{\tilde{m}}(\tilde{k})$ and
$\tilde{h}^{\tilde{k}}_{\tilde{m}}$ in the same way.

Note that the atoms with $u=0$ and $u=1$ contribute only to the
diagonal and off diagonal terms of $H_m(k)$, respectively.
Consequently, in the dispersion relations \eqref{EeboovalC}, the
interactions of C$_{000}$ with C$_{ts0}$ atoms determine for each $k$
the average energy of two bands, while the interactions with
C$_{ts1}$ atoms shifts up and down symmetrically this average to the
eigen energies. Within $\pi^\bot$-orbitals tight-binding
approximation this result is obviously not restricted to some of the
neighbors and includes the local distortions induced by the
cylindrical geometry. Further, note that $H_{tsu,t's'u'}$ would be
equal to $\bra{tsu}H\ket{t's'u'}$ if and only if the atomic orbitals
$\ket{tsu}$ were orthonormal basis. Since expressed in terms of
$H_{tsu,t's'u'}$ matrix elements, equations \eqref{EebooC} refer to
the realistic nonorthonormal case (therefore the resulting Bloch
functions are not orthonormalized). To take advantage of the
calculated~\cite{SEIFERT} elements $\bra{tsu}H\ket{t's'u'}$ and the
overlap integrals $\braket{000}{tsu}$ one uses another matrix having
also the form \eqref{EH|0C}, but with $\bra{000}H\ket{tsu}$ instead
of $H_{tsu}$. Multiplying \eqref{EH|0Ca}  by the inverse of the
analogous matrix of the overlap integrals, one gets $H_m(k)$
completely in terms of the known Slater-Koster elements and the
overlap integrals.

Also, the resultgiven in \eqref{EeboovalC} is general in the sense
that the eigen energies at the edges of the irreducible domain can be
obtained from this expression by substituting the limiting values of
$km$ (or $\tilde{k}\tilde{m}$) numbers; the nondegenerate ones point
out the states with $U$-parity. For the $\cZ$ and $\cA$ tubes the
dispersion relations can be derived, too: only $n_1=q/2=n$ and
$n_2=0$ for the zig-zag or $n_1=n_2=n$ for the armchair tubes should
be used. In these cases \eqref{EeboovalC} is the same for $m$ and
$-m$, reflecting the anticipated general conclusion that the bands of
the achiral tubes are four-fold degenerate apart from the double
degenerate $m=0,n$ bands. These double degenerate bands are with even
$\sigma_{\mathrm{v}}$-parity for $\cZ$ tubes, while they form two
pairs with opposite $\sigma_{\mathrm{v}}$ parity in $\cA$ tubes.
Nevertheless, the symmetry adapted basis for the four-fold degenerate
bands and the representations with parities cannot be such
straightforwardly derived from \eqref{EeboovecC}, and each of these
representation should be considered separately.

As for the most usual orthogonal orbitals nearest neighbors
approximation, the sums in \eqref{EH|0C} are restricted to the
constant term $H_{000}$ in $h^0_m(k)$ and to the three nearest
neighbors in $h^1_m(k)$. Taking $H_{000}=0$, i.e. shifting the energy
scale for $H_{000}$, and substituting in \eqref{Ephase} for the
nearest neighbors (Fig. \ref{Fswcneigh}) the parameters
\begin{eqnarray}\label{Eneigh}
 t_1=-\frac{n_2}{n},&\quad& s_1=\frac{2n_1+(1+r\cR)n_2}{q\cR},\nonumber\\
 t_2=\frac{n_1}{n},&\quad& s_2=\frac{(1-r\cR)n_1+2n_2}{q\cR},\nonumber\\
 t_3=t_1+t_2,&\quad& s_3=s_1+s_2,\nonumber
\end{eqnarray}
one gets pairs of equally assigned double degenerate generic bands
for each allowed $m$ (i.e. $\tilde{m}$):
\begin{equation}\label{EbandC}
\epsilon^\pm_{E_m}(k)=\pm|\sum_{i=1}^3H_{t_is_i1}\re^{\ri
\psi^k_m(t_i,s_i)}|.
\end{equation}
As usual, the same expression but with ${\psi}^{\tk}_{\tm}(t_i,s_i)$
gives the bands $\epsilon^\pm_{E_{\tilde{m}}}(\tilde{k})$ assigned by
$\tilde{k}\tilde{m}$-numbers.

\begin{widetext}\begin{center}\begin{table}[h]\centering
\caption{\label{TbandsgBf} {\bf Bands and symmetry-adapted eigen
vectors of the carbon nanotubes.} For each irreducible representation
the corresponding frequency number $N$, energy $\epsilon$ in the
simplest (orthogonal orbitals, nearest neighbors, homogeneous
distortions) model and generalized Bloch functions $\ket{km\Pi}$ of
the corresponding bands are given. Substituting $\psi^k_m$ and
$h^k_m$ by $\tilde{\psi}^{\tilde{k}}_{\tilde{m}}$ and
$\tilde{h}^{\tilde{k}}_{\tilde{m}}$ one obtains these quantities in
$\tilde{k}\tilde{m}$ numbers (see \eqref{Ephase} and \eqref{EH|0C}
and the comment bellow).
$\gamma=\mathrm{Arg}(1+2\re^{\ri\frac{ka}{2}}\cos\frac{\pi m}{n})$.}
\label{Tbands}\begin{tabular}{@{}llll}\toprule
$\cC$&$N$&$\epsilon$&{Generalized Bloch functions}\\ \colrule
${_0}A^\Pi_m$&$1$&$V\Pi(1+2\re^{2\ri\frac{m\pi}{q}})$&
$\ket{0m\Pi}=\frac{1}{\sqrt{|{\bm L_\cC}|}}\sum_{ts}
  \re^{-\ri\psi^0_m(t,s)}(\ket{ts0}+\Pi\ket{ts1})$\\
${_\pi}A^\Pi_m$&$1$&$-V\Pi$& $\ket{\pi m\Pi}=\frac{1}{\sqrt{|{\bm
L_\cC}|}} \sum_{ts}
  \re^{-\ri\psi^{\pi}_m(t,s)}(\ket{ts0}+\Pi\ket{ts1})$\\
${_k}E_m$&$2$& $\pm|V\sum_i \re^{\ri\psi^k_m(t_i,s_i)}|$ &
$\ket{km;\pm}=\frac{1}{\sqrt{|{\bm L}_\cC|}} \sum_{ts}
  \re^{-\ri\psi^k_m(t,s)}(\ket{ts0}
  \pm\re^{\ri h^k_m}\ket{ts1})$\\
  & & & $\ket{-k,-m;\pm}=\frac{1}{\sqrt{|{\bm L}_\cC|}}\sum_{ts}
  \re^{\ri\psi^k_m(t,s)}(\ket{ts1}
  \pm\re^{\ri h^k_m}\ket{ts0})$\\ \toprule
$\cZ$&$N$&$\epsilon$&{Generalized Bloch functions}\\ \colrule
${_0}A^\Pi_m$&1&$V\Pi(1+2\re^{\ri\frac{m\pi}{n}})$& $\ket{0m\Pi
A}=\sqrt{\frac{2}{|{\bm L_{\cZ}}|}}\sum_{ts}
  \re^{-\ri\frac{m\pi}{n}t}(\ket{ts0}+\Pi\ket{ts1})$\\
${_0}E^\Pi_m$&$1$&$V\Pi(1+2\cos\frac{m\pi}{n})$&
$\ket{0m\Pi}=\sqrt{\frac{2}{|{\bm L_{\cZ}}|}}\sum_{ts}
  \re^{-\ri\frac{m\pi}{n}(2s+t)}(\ket{ts0}+\Pi\re^{\ri\frac{2m\pi}{n}}\ket{ts1})$\\
  & & & $\ket{0,-m,\Pi}=\sqrt{\frac{2}{|{\bm L_{\cZ}}|}}\sum_{ts}
   \re^{\ri\frac{m\pi}{n}(2s+t)}(\re^{\ri\frac{2m\pi}{n}}\ket{ts0}+\Pi\ket{ts1})$\\
${_k}E^A_m$&$2$&$\pm|V|\sqrt{5+4\re^{\ri\frac{m\pi}{n}}\cos\frac{ka}{2}}$&
$\ket{kmA;\pm}=\sqrt{\frac{2}{|{\bm L_{\cZ}}|}}\sum_{ts}
  \re^{-\ri(\frac{m\pi}{n}+\frac{ka}{2})t}(\ket{ts0}
  \pm\re^{\ri h^m_k}\ket{ts1})$\\
  & & & $\ket{-k,m,A;\pm}=\sqrt{\frac{2}{|{\bm L_{\cZ}}|}} \sum_{ts}
   \re^{-\ri(\frac{m\pi}{n}-\frac{ka}{2})t}(\ket{ts1}
   \pm\re^{\ri h^m_k}\ket{ts0})$\\
${_\pi}E^\Pi_\frac{n}{2}$&$1$&$-V\Pi$&$\ket{\pi,\frac{n}{2},\Pi}=\sqrt{\frac{2}{|{\bm
 L_{\cZ}}|}}
  \sum_{ts}(-1)^{s+t}(\ket{ts0}+\Pi\ket{ts1})$\\
  & & & $\ket{\pi,-\frac{n}{2},\Pi}=\sqrt{\frac{2}{|{\bm L_{\cZ}}|}}
   \sum_{ts}(-1)^s(\ket{ts0}+\Pi\ket{ts1})$\\
${_k}G_m$&$2$&$\pm
  V\sqrt{1+4\cos\frac{ka}{2}\cos\frac{m\pi}{n}+4\cos^2\frac{m\pi}{n}}$& $\ket{km;\pm}=\sqrt{\frac{2}{|{\bm L_{\cZ}}|}}
\sum_{ts}
  \re^{-\ri\frac{ka}{2}t}\re^{-\ri\frac{m\pi}{n}(t+2s)}(\ket{ts0}
  \pm\re^{\ri\gamma}\re^{\ri\frac{2m\pi}{n}}\ket{ts1})$\\
  & & &$\ket{k,-m;\pm}=\sqrt{\frac{2}{|{\bm L_{\cZ}}|}} \sum_{ts}
   \re^{-\ri\frac{ka}{2}t}\re^{\ri\frac{m\pi}{n}(t+2s)}(\re^{\ri\frac{2m\pi}{n}}\ket{ts0}
   \pm\re^{\ri\gamma}\ket{ts1})$\\
  & & &$\ket{-k,m;\pm}=\sqrt{\frac{2}{|{\bm L_{\cZ}}|}} \sum_{ts}
   \re^{\ri\frac{ka}{2}t}\re^{-\ri\frac{m\pi}{n}(t+2s)}(\re^{\ri\frac{2m\pi}{n}}\ket{ts1}
   \pm\re^{\ri\gamma}\ket{ts0})$\\
  & & &$\ket{-k,-m;\pm}=\sqrt{\frac{2}{|{\bm L_{\cZ}}|}} \sum_{ts}
   \re^{\ri\frac{ka}{2}t}\re^{\ri\frac{m\pi}{n}(t+2s)}(\ket{ts1}
   \pm\re^{\ri\gamma}\re^{\ri\frac{2m\pi}{n}}\ket{ts0})$\\ \toprule
$\cA$&$N$&$\epsilon$&{Generalized Bloch functions}\\ \colrule
${_0}\Pi^+_m$&1&$V\Pi(1+2\re^{\ri\frac{m\pi}{n}})$&
$\ket{0m+\Pi}=\sqrt{\frac{2}{|{\bm L_{\cA}}|}} \sum_{ts}
  \re^{-\ri\frac{m\pi}{n}t}(\ket{ts0}+\Pi\ket{ts1})$\\
${_0}E^+_m$&$2$&$\pm|V|\sqrt{5+4\cos\frac{m\pi}{n}}$&$\ket{0m+;\pm}=\sqrt{\frac{2}{|{\bm
L_{\cA}}|}} \sum_{ts}
  \re^{-\ri\frac{m\pi}{n}(2s+t)}(\ket{ts0}\pm\re^{\ri h^m_0}\ket{ts1})$\\
  & & & $\ket{0,-m,+;\pm}=\sqrt{\frac{2}{|{\bm L_{\cA}}|}} \sum_{ts}
   \re^{\ri\frac{m\pi}{n}(2s+t)}(\ket{ts1}\pm\re^{\ri h^m_0}\ket{ts0})$\\
${_k}E^\Pi_m$&$1$&$V\Pi(1+2\re^{\ri\frac{m\pi}{n}}\cos\frac{ka}{2})$&
$\ket{km\Pi}=\sqrt{\frac{2}{|{\bm L_{\cA}}|}} \sum_{ts}
  \re^{-\ri(\frac{m\pi}{n}+\frac{ka}{2})t}(\ket{ts0}+\Pi\ket{ts1})$\\
  & & &$\ket{-k,m,\Pi}=\sqrt{\frac{2}{|{\bm L_{\cA}}|}} \sum_{ts}
  \re^{-\ri(\frac{m\pi}{n}-\frac{ka}{2})t}(\ket{ts0}+\Pi\ket{ts1})$\\
${_\pi}E^\Pi_\frac{n}{2}$&$1$&$-V\Pi$&$\ket{\pi,\frac{n}{2},\Pi}=\sqrt{\frac{2}{|{\bm
L_{\cA}}|}}\sum_{ts}
  (-1)^{s+t}(\ket{ts0}+\Pi\ket{ts1})$\\
  & & &$\ket{\pi,-\frac{n}{2},\Pi}=\sqrt{\frac{2}{|{\bm L_{\cA}}|}}\sum_{ts}
   (-1)^s(\Pi\ket{ts0}+\ket{ts1})$\\
${_k}G_m$&$2$&$\pm V\sqrt{1+4\cos\frac{ka}{2}\cos\frac{m\pi}{n}
 +4\cos^2\frac{ka}{2}}$&$\ket{km;\pm}=\sqrt{\frac{2}{|{\bm L_{\cA}}|}}
\sum_{ts}
  \re^{-\ri\frac{ka}{2}t}\re^{-\ri\frac{m\pi}{n}(t+2s)}(\ket{ts0}
  \pm\re^{\ri h^m_k}\ket{ts1})$\\ & & & $\ket{k,-m;\pm}=\sqrt{\frac{2}{|{\bm L_{\cA}}|}}\sum_{ts}
   \re^{-\ri\frac{ka}{2}t}\re^{\ri\frac{m\pi}{n}(t+2s)}(\ket{ts1}
   \pm\re^{\ri h^m_k}\ket{ts0})$\\ & & &$\ket{-k,m;\pm}=\sqrt{\frac{2}{|{\bm L_{\cA}}|}}\sum_{ts}
   \re^{\ri\frac{ka}{2}t}\re^{-\ri\frac{m\pi}{n}(t+2s)}(\ket{ts0}
   \pm\re^{\ri h^m_k}\ket{ts1})$\\ & &
&$\ket{-k,-m;\pm}=\sqrt{\frac{2}{|{\bm L_{\cA}}|}}\sum_{ts}
   \re^{\ri\frac{ka}{2}t}\re^{\ri\frac{m\pi}{n}(t+2s)}(\ket{ts1}
   \pm\re^{\ri h^m_k}\ket{ts0})$
\\ \botrule
\end{tabular}\end{table}\end{center}\end{widetext}

\clearpage

Finally, the rolling up induced differences in the interatomic
distances of the honeycomb lattice are frequently neglected
(homogeneous distortions approximation), which is achieved by setting
$H_{t_is_i 1}=V$ for the nearest neighbors ($V$ is estimated between
-3.003~eV and -2.5~eV). All the dispersion relations and the
corresponding eigen states (in the form of generalized Bloch sums)
are given in the Table \ref{TbandsgBf} for this approximation, and in
Fig. \ref{Fcband} the assignation of these bands for several tubes is
presented.

\section{Selection rules}\label{Sselrul}
One of the most important benefits from the assignation by all
quantum numbers comes through the applications of the selection rules
in various calculations of physical properties of nanotubes. As it
was shown in the section \ref{Sqn}, each allowed pair of numbers
$(k,m)$ or $(\tilde{k},\tilde{m})$, together with the parities when
necessary, singles out states corresponding to one irreducible
representation. In this sense, a representation is specified by
$(km\Pi)$, where $\Pi$ stands for all possible parities. If the
representation is degenerate the "raw" index $r$ running from 1 to
the dimension of the representation is used to enumerate the states
of the same irreducible subspace (with the same eigen energy).
Altogether, the state is denoted as $\ket{km\Pi r;F}$. For example,
the symmetry adapted eigen states of the ${_k}E_m$ electronic bands
of $\cC$-tubes (section \ref{Sbands}) are now denoted as
$\ket{km1;\pm}=\ket{km;\pm}$ and $\ket{km2;\pm}=\ket{-k,-m;\pm}$.
Also the components  $Q^{(km\Pi)}_r$ of the physical tensor $Q$ are
associated to the same quantum numbers (i.e. irreducible
representations), giving their transformation rules under the line
group symmetry operations. Then, the matrix elements of $Q$ are
expressed in the Wigner-Eckart form~\cite{MESSIAH}:
\begin{widetext}\begin{equation}\label{EWE}
\bra{k_fm_f\Pi_fr_f;F_f}Q^{(km\Pi)}_r\ket{k_im_i\Pi_ir_i;F_i}=
\braket{k_fm_f\Pi_fr_f}{km\Pi,r;k_im_i\Pi_ir_i}
Q(k_fm_f\Pi_f;F_f\Vert km\Pi\Vert k_im_i\Pi_i;F_i).
\end{equation}\end{widetext}
Here,  $Q(k_fm_f\Pi_f;F_f\Vert km\Pi\Vert k_im_i\Pi_i;F_i)$, the
reduced matrix element, is independent on the indices $r$, $r_i$ and
$r_f$. The Clebsch-Gordan coefficients $\braket{k_fm_f\Pi_fr_f}{km\Pi
r;k_im_i\Pi_ir_i}$, being independent on $Q$, are {\sl a priori}
given by the symmetry of the system; the matrix elements are thus
subjected to the selection rules showing when these coefficients are
nonzero.

The Clebsch-Gordan coefficients comprise complete information on the
selection rules. For the SWCT symmetry groups \eqref{ELG} they are
given in the appendix. Generally they reflect conservation laws of
the linear momentum $\Delta k=k_f-k_i\dot=k$, helical momentum
$\Delta \tilde{k}=\tilde{k}_f-\tilde{k}_i\dot=\tilde{k}$, pure
angular momentum
$\Delta\tilde{m}=\tilde{m}_f-\tilde{m}_i\dot=\tilde{m}$ and parities
$\Pi_f=\Pi\Pi_i$ (assuming $+1$ for "$+$" or $A$, and $-1$ for "$-$"
and $B$). The $z$ component of the total quasi angular momentum is
not conserved since $\Delta m=m_f-m_i\dot=m+Kp$ and $Kp$ may not be a
multiple of $q$. In fact, among the isogonal rotations $C^s_q$ only
the $C^s_n$ are symmetries of nanotubes, and the corresponding
$\tilde{m}$ is conserved, since $p$ is a multiple of $n$. This is
important in the Umklapp processes, when $K$ is non-vanishing and $m$
is changed.

The symmetry properties of the most interesting tensors are
expressed~\cite{WOOSTER} in terms of the three dimensional vector
representations $D^p$ and $D^a$ (polar and axial) of $\bL$, since
these tensors are functions of the radius vector ${\bm r}$, momentum
${\bm p}$, electrical field ${\bm E}$ (polar vectors), angular
momentum ${\bm l}$ and magnetic field ${\bm H}$ (axial vectors). The
irreducible components of the corresponding representations are given
in the Table \ref{Ttensors}. For all of them $k=0$, causing that only
direct processes are encountered and now $m$ is also a conserved
quantum number (since $k_i=k_f$ yields $K=0$). This means that their
symmetry properties are related to the isogonal groups
(\ref{Eisogroups}).

To facilitate the application of \eqref{EWE} we discuss the general
forms of some of the tensors indicated in a Table \ref{Ttensors}
being related to the optical properties~\cite{LANDAU} of nanotubes.

In the linear approximation  the tensor of the dielectric
permeability in the weak external electric field ${\bm E}$ is
$\varepsilon_{[ij]}({\bm E})=\varepsilon_{[ij]}(0)+
\sum_k\alpha_{[ij]k}{\bm E}_k$. For the chiral tubes, the general
form of the zero field permeability tensor~\cite{YITR} is
$\varepsilon(0)=\mathrm{diag}(\varepsilon_{xx},\varepsilon_{xx},
\varepsilon_{zz})$. As the frequency number of the trivial
representation ${_0}A^+_0$ in the $\alpha_{[ij]k}$ is equal to one,
the single parameter $\alpha$, determined by the tube microscopic
properties, controls the field-dependent dielectric permeability
behaviour: $\varepsilon({\bm
E})=\Big(\begin{smallmatrix}\varepsilon_{xx}&0&\alpha E_y\cr
0&\varepsilon_{xx}&-\alpha E_x\cr \alpha E_y&-\alpha
E_x&\varepsilon_{zz}\cr\end{smallmatrix}\Big)$. Thus, the optical
activity of $\cC$ tubes is changed by the perpendicular electric
field, and instead of one there are two optic axes whose direction
depend on the applied field. For the $\cZ$ and $\cA$ tubes the
external field does not change their optical {\em symmetry}, since no
trivial component appears in the decomposition.

The electromagnetic response to a weak applied field is characterized
by the dielectric function $\varepsilon_{ij}({\bm k},\omega)$.
Although optical absorption and diffraction are well described within
the long-wavelength limit, for the optical activity the terms of
$\varepsilon_{ij}$ linear in the components of the wave vector ${\bm
k}$ (having different symmetry from the ${\bm k}$-independent ones)
should be considered. These linear terms define the tensor
$\gamma_{ijk}$ and its symmetric and antisymmetric~\cite{TASAKI}
parts with respect to the last two subscripts:
$\gamma_{ijk}=\left[\partial\varepsilon_{ij}({\bm k},\omega
)/\partial k_l\right]_{{\bm k}=0}=\ri
(\gamma^A_{i\{jl\}}+\gamma^S_{i[jl]})$. For the $\cZ$ and $\cA$ tubes
there is no linear optical response while for the chiral tubes the
antisymmetric part $\gamma_{i\{jl\}}^A$ is determined by two
independent parameters involved in six nonvanishing tensor elements:
$\gamma_{xyz}^A=\gamma_{yzx}^A=-\gamma_{xzy}^A=-\gamma_{yxz}^A$ are
related to the interband transitions
${_0}A_0^\pm{\,\leftrightarrow\,}{_0}E_1$,
${_k}E_m{\,\leftrightarrow\,}{_k}E_{m+1}\ (k\in [0,\pi ])$,
${_0}A_{q/2}{\,\leftrightarrow\,}{_0}E_{q/2-1}$ (this follows from
(\ref{EWE}) when the operators $p_x$, $p_y$, $l_x$ and $l_y$ are
substituted for $Q$), while $\gamma_{zxy}^A=-\gamma_{zyx}^A$ are
related to the interband transitions
${_0}A_0^\pm{\,\leftrightarrow\,}{_0}A_0^\mp$ and
${_0}A_{q/2}^\pm{\,\leftrightarrow\,}{_0}A_{q/2}^\mp$ (now $p_z$ and
$l_z$ are used). The single independent parameter of
$\gamma^S_{i[jl]}$ is involved in the four nonvanishing tensor
elements:
$\gamma_{xyz}^S=\gamma_{xzy}^S=-\gamma_{yxz}^S=-\gamma_{yzx}^S$
related to the interband transitions induced by the symmetric
operator $\frac{1}{2}({\hat z}{\hat p}_y+{\hat y}{\hat p}_z+H.c.)$.

The conductivity tensor $\sigma_{ij}$ of a system in a sufficiently
weak magnetic field ${\bm H}$ is well approximated quadratically:
$$\sigma_{ij}({\bm
H})=\sigma_{[ij]}(0)+\sum_{k=1}^3\rho_{\{ij\}k}H_k+
\sum_{k=1}^3\sum_{l=1}^3\beta_{[ij][kl]}H_kH_l,$$
where the symmetry~\cite{YITR,Y-I} allows a symmetric tensor
$\sigma_{[ij]}(0)=\mathrm{diag}(\sigma_{xx},\sigma_{xx},\sigma_{zz})$.
The 3rd rank tensor $\rho_{\{ij\}k}$ is responsible for the linear
contribution of the field (Hall effect), while the 4th rank tensor
$\beta_{[ij][kl]}$ introduces a small correction to the main effect.
Because of the symmetry, $\rho_{\{ij\}k}$  is of the same form for
all SWCT ($\cC,\cZ,\cA$): two independent parameters $\rho_1$ and
$\rho_2$ define its six nonvanishing tensor components:
$\rho_{xyz}=-\rho_{yxz}=\rho_1$,
$\rho_{xzy}=-\rho_{zxy}=\rho_{zyx}=-\rho_{zxy}=\rho_2$. Also
$\beta_{[ij][kl]}$ is of the same form for the chiral and the achiral
SWCT, with six independent parameters involved within altogether 21
nonzero components:
$\beta_{xxxx}=\beta_{yyyy}=\beta_1$, $\beta_{zzzz}=\beta_2$,
$\beta_{xxyy}=\beta_{yyxx}=\beta_3$
$\beta_{xyxy}=\beta_{xyyx}=\beta_{yxxy}=\beta_{yxyx}=
  \frac{1}{2}(\beta_1-\beta_3)$,
$\beta_{xxzz}=\beta_{yyzz}=\beta_4$,
$\beta_{xzxz}=\beta_{xzzx}=\beta_{yzyz}=\beta_{yzzy}=
 \beta_{zxxz}=\beta_{zxzx}=\beta_{zyyz}=\beta_{zyzy}=\beta_5$,
$\beta_{zzxx}=\beta_{zzyy}=\beta_6$.
So, the conductivity tensor $\sigma$ of SWCT in the presence of
the weak magnetic field $\bH$, up to the square terms in the
applied field, is of the form:
\begin{widetext}
$$\sigma({\bm H})=\sigma+\begin{pmatrix}
 \beta_1H_x^2+\beta_3H_y^2+\beta_4H_z^2&
  \rho_1H_z+(\beta_1-\beta_3)H_xH_y&
  \rho_2 H_y+2\beta_5H_xH_z\cr
 -\rho_1H_z+(\beta_1-\beta_3)H_xH_y&
  \beta_3H_x^2+\beta_1H_y^2+\beta_4H_z^2&
  -\rho_2H_x+2\beta_5H_yH_z\cr
 -\rho_2H_y+2\beta_5H_xH_z&
  \rho_2H_x+2\beta_5H_yH_z&
  \beta_6(H_x^2+H_y^2)+\beta_2H_z^2\cr\end{pmatrix}.$$

\begin{table}[hbt]\caption[]{\label{Ttensors}
{\bf Symmetry of the tensors of SWCT.} The decompositions onto
irreducible representations of the most frequent tensors (given in
the last column) of the chiral (column 2) and the zig-zag and the
armchair (column 3) SWCT. Tensors are obtained by multiplying polar
and axial vectors, and the type of the products ($\otimes$, $[\dots]$
and $\{\dots\}$ for the direct, symmetrized and antisymmetrized) is
in the first column. For $\cC$ tubes $\kappa=0$ for the $km$ numbers
and $\kappa=2\pi r/qa$ for the $\tilde{k}\tilde{m}$ numbers.}
\begin{tabular}{llll}\toprule
Type&$\cC$ tubes&$\cZ$ and $\cA$ tubes&Tensor\\ \colrule
$D^{p}$&${_0}A^-_0+{_\kappa}E_1$&$ {_0}A^-_0+{_0}E^+_1$&$
r_i,p_i,E_i$\\
$D^{a}=\{D^{a/p^2}\}$&${_0}A^-_0+{_\kappa}E_1$&$
{_0}B^+_0+{_0}E^-_1$&$ l_i,H_i,R_{\{ij\}}$\\
$D^{a/p^2}$&$2{_0}A^+_0+{_0}A^-_0+2{_\kappa}E_1+{_{2\kappa}}E_2$&$
2{_0}A^+_0+{_0}B^+_0+2{_0}E^-_1+{_0}E^+_2$&$ \rho_{\{ij\}k},R_{ij}$\\
$[D^{a/p^2}]$&$2{_0}A^+_0+{_\kappa}E_1+{_{2\kappa}}E_2$&$
2{_0}A^+_0+{_0}E^-_1+{_0}E^+_2$&$
\varepsilon_{[ij]},\sigma_{[ij]},R_{[ij]}$\\
$D^{p}\otimes
D^{a}$&$2{_0}A^+_0+{_0}A^-_0+2{_\kappa}E_1+{_{2\kappa}}E_2$&$
{_0}A^-_0+2{_0}B^-_0+2{_0}E^+_1+{_0}E^-_2$&$\gamma^A_{i\{jk\}}$\\
$D^{p}\otimes[D^{a/p^2}]$&$
{_0}A^+_0+3{_0}A^-_0+4{_\kappa}E_1+2{_{2\kappa}}E_2+{_{3\kappa}}E_3$&$
3{_0}A^-_0+{_0}B^-_0+4{_0}E^+_1+2{_0}E^-_2+{_0}E^+_3$&$
\alpha_{[ij]k},\gamma^S_{i[jl]}$\\
$D^{p^3}$&$
3{_0A}^+_0+4{_0}A^-_0+6{_\kappa}E_1+3{_{2\kappa}}E_2+{_{3\kappa}}E_3$&$
4{_0}A^-_0+3{_0}B^-_0+6{_0}E^+_1+3{_0}E^-_2+{_0}E^+_3$&$
\gamma_{ijk}$\\
$[D^{a/p^2}]\otimes[D^{a/p^2}]$&$
6{_0}A^+_0+2{_0}A^-_0+6{_\kappa}E_1+5{_{2\kappa}}E_2+2{_{3\kappa}}E_3+{_{4\kappa}}E_4$&$
6{_0}A^+_0+2{_0}B^+_0+6{_0}E^-_1+5{_0}E^+_2+2{_0}E^-_3+{_0}E^+_4$&
$\beta_{[ij][kl]}$\\ \botrule
\end{tabular}\end{table}
\end{widetext}

In general, the Raman (polarizability) tensor $R_{ij}$ relates
induced polarization to the external electric field~\cite{RAMAN}:
$P_i=\sum_jR_{ij}E_j$. Therefore, the Table \ref{Ttensors} combined
with \eqref{ECGgen} gives the selection rules of the Raman
scattering: the relevant transitions are between the states with
$km$-numbers $k_f-k_i=0$ and $\Delta m=m_f-m_i=0,\pm1,\pm2$; for the
achiral tubes $z$-reversal parity of these states is different if
$\Delta m=1$ and same if $\Delta m$ even. For the frequently
important symmetric part $R_{[ij]}$ and its anisotropic component
$R^{\mathrm{a}}_{[ij]}$ (the last one transforms according to
$[D^{p^2}]-{_0A}^+_0$), the momenta selection rules are same, while
both the $z$-reversal (and the vertical mirror for achiral tubes)
parity is conserved if $\Delta m=0$. The isotropic component
$R^s_{[ij]}$ transforms according to the identity representation
${_0A}^+_0$, and involves only the transitions between the states
with the coincident quantum numbers. As for the antisymmetric part,
$R_{\{ij\}}$, $\Delta m=0,\pm1$; if $\Delta m=0$ the relevant
transitions are between the states with opposite $U$-parity for the
chiral tubes and equal horizontal mirror parity but the opposite
vertical mirror parity in the achiral cases. Of course, in the
concrete calculations, these rules can be further specified according
to the incident light polarization and direction.

\section{Concluding remarks}
The assignation of the energy bands of SWCT by the complete set of
the symmetry based quantum numbers is discussed. The quantum numbers
$k$, $m$ (or $\tilde{k}$, $\tilde{m}$) and parities $\pm$ and $A/B$
are related to specific symmetries of SWCT. When parametrized by the
quasi momentum $k$, the bands carry the quantum number of the angular
momentum $m$, while in the parametrization by the helical momentum
$\tk$ only the uncoupled to the translations part $\tm$ of the
angular momentum characterizes the bands. The ranges of $m$ and $\tm$
have been redefined compared to the one used in the nanotube
literature~\cite{ELBANDS-I,ELBANDS-Y} to get the standard quantum
mechanical interpretation of the $z$-projection of the orbital
angular momentum. The momenta quantum numbers are imposed by the
roto-translational subgroup $\bL^{(1)}=\bT^q_r{\bm C}_n$, and
characterize all the quasi 1D crystals. Indeed, in their symmetry
$\bL^{(1)}$ is always present as the maximal abelian part, thus
causing no degeneracy. Additional $U$ and $\sigma_{\mathrm{v}}$
parities of SWCT introduce band degeneracy. Relating the quantum
numbers to the irreducible representations of the symmetry groups
this assignation immediately gives the band degeneracies and
information on non-accidental band sticking.

The bands have specific symmetry with respect to the $k=0$ and
$k=\pi/a$; therefore, the irreducible domain sufficient to
characterize the entire band is the non-negative half of the
Brillouin zone $[0,\pi]$. At the edge points $k=0,\pi/a$, either the
band stick to the another one or the corresponding eigen state gets
$z$-reversal parity $\pm$. The $U$-axis symmetry reverses both the
linear and angular momenta causing at least double degeneracy of the
bands in the interior of the ID. For the $\cZ$ and $\cA$ tubes, the
vertical mirror plane implies degeneracy of $m$ and $-m$ bands. Thus
the bands are four-fold degenerate, except $m=0,n$ ones, which are
$\sigma_{\mathrm{v}}$ odd or even and double degenerate.

At $k=0$ $m$ and $-m$ bands are sticked together. The symmetry of the
screw axis imposes additional band sticking: the set of $q/n$ bands
is pairwise ($m$ and $-p-m$ at $k=\pi/a$ and $\pm m$ at $k=0$)
sticked together. These bands are continued in a single band in the
$q/n$ times extended Brillouin zone corresponding to the
$\tilde{k}\tilde{m}$ quantum numbers. Only the bands ending up with
$U$-parity even or odd states are not sticked to the another ones,
with the van Hove singularities and the halved degeneracy at the end
points.

These conclusions are generally valid for any (quasi)particle energy
bands of SWCT. All other bands sticking or increased degeneracy, if
any, are accidental, i.e. related to the hamiltonian under study.
Note that only within spin independent models the $U$-axis imposed
double degeneracy coincides with that introduced by the time reversal
symmetry, since both operations reverse linear and orbital angular
momenta.

According to this general scheme the complete assignation of the SWCT
electronic tight-binding bands is performed. The generalized Bloch
functions are found and characterized by the full set of $(km\Pi)$,
or, alternatively, $(\tilde{k}\tilde{m}\Pi)$, quantum numbers. All
these functions contain two parts: the two halves of SWCT consisting
of C$_{ts0}$ and C$_{ts1}$ atoms (black and white ones in the Fig.
\ref{Fswcneigh}) contribute to the state by different phase factors.
This form is useful in calculations and comparison to the STM
images~\cite{STM}, again manifests the existence of the $U$-symmetry
which interrelate the two halves.

A brief comment on the SWCT conductivity within the present context
may enlighten some of the discussed questions. Recall that the
simplest (tight-binding nearest neighbors and homogeneous
distortions) model with the bands given in the Table \ref{TbandsgBf},
predicts~\cite{ELBANDS-I,ELBANDS-Y} that the tubes with $n_1-n_2$
divisible by 3 should be conductors due to the  crossing of the two
bands~\cite{YG2} $\tilde{m}_F=0$ if $\cR=3$ and
$\tilde{m}_F=(-1)^{\Fr{(r/3)}}n/3$ for $\cR=1$ at
$\tilde{k}_F=2\pi/3$. Alternatively, in $km$ numbers, when $\cR=3$
then $k_F=2\pi/3$ and $m_F=nr \pmod{q}$, while $k_F=0$ and $m_F=\pm
q/3$ for $\cR=1$. This extra degeneracy at the Fermi level is a model
dependent accidental one, being not induced by symmetry. On the
contrary, the symmetry based non-crossing rule just prevents the
conductivity except in the armchair tubes, since the momenta quantum
numbers of the crossing bands are the same; only for the armchair
tubes, when $m_F=n$ these bands also carry the opposite vertical
mirror parity. So, as verified experimentally~\cite{ZHO00} and in the
more subtle theoretical models~\cite{YG2}, the secondary gap must be
opened except for the armchair tubes, for which the accidental
crossing point $k_F$ is shifted to the left. In these cases the
metallic plato~\cite{STM,PLATO} is ended by the systematic van Hove
singularities.

The major benefit from the complete assignation of bands and
corresponding generalized Bloch functions comes from the selection
rules. The momenta conservation selection rules \eqref{Eselrule}
emerge from the roto-translational subgroup $\bT^r_q{\bm C}_n$ making
these rules also applicable to all other nanotubes (multi-wall, BN,
etc.) and stereoregular polymers. The novel conserved parities refine
the momenta conservation rules. The coincidence of the $z$-reversal
odd and even states with the systematic van Hove singularities proves
substantial influence of the parities to the physical processes in
nanotubes and related spectra~\cite{SPECTRA,TASAKI}. Therefore, these
additional rules must not be overlooked in calculations.

To illustrate further the relevance of the derived parity selection
rules, let us briefly discuss armchair tubes and the parallel
component of the dielectric tensor $\epsilon_{ij}({\bm k},\omega)$,
which is the corner stone in the analysis of various optical
properties~\cite{TASAKI}. The contribution of the direct interband
transitions caused by the electric field along the $z$-axis are to be
included in calculations. As the perturbation field has odd
$z$-reversal and even vertical mirror parities, it transforms
according to the representation ${_0}A^-_0$. Therefore, the
absorption may be realized only by the (vertical) transitions
$\epsilon^-_m(k)\rightarrow\epsilon^+_m(k)$, and this exhausts the
selection rules imposed by the roto-translational subgroup.
Nevertheless, the eigen states of the pairs of the double degenerate
bands with $m=0,n$ have different $\sigma_{\mathrm{v}}$ parity, and
the transitions between these bands are forbidden for any $k$. Thus,
only the transitions between the four fold degenerate ${_k}G_m$ bands
are allowed for $z$ polarized light. Also in the Raman scattering
processes the selection rules besides the momenta strongly involve
parities.

Finally, the tensor properties of some physical quantities were
established, to make the use of the selection rules quite
straightforward. We emphasize that the considered tensors interrelate
vector (polar or axial) quantities making that all of them are
associated to quantum number $k=0$. This provides full conservation
of momenta (e.g. vertical optical transitions), even when the
$km$-numbers are used (where $m$ is not conserved in general). On the
other side, the selection rules with the conserved
$\tilde{k}\tilde{m}$ numbers are more compact and easy to deal with.
Nevertheless, some components of the relevant tensors have
non-vanishing $\tilde{k}$, which makes some results less obvious. For
example, some optical transitions are not vertical in $\tilde{k}$.

\appendix
\section{Clebsch-Gordan coefficients}\label{Acgcoef}
The Clebsch-Gordan coefficients are given for the irreducible
representations of the line groups $\bL_\cC$ and $\bL_\cA$ presented
in Ref. \onlinecite{YTBA}. Besides the pair $(k,m)$ of linear and
$z$-component of the angular momenta (or $(\tilde{k},\tilde{m})$ of
helical and pure angular momenta), some of the irreducible
representations carry also the quantum numbers of parities with
respect to the $U$ axis, and mirror planes $\sigma_{\mathrm{v}}$ or
$\sigma_{\mathrm{h}}$. Thus the following Clebsch-Gordan coefficients
reflect the conservation laws of these quantities.

The addition of quasi momenta is performed modulo their range, which
is indicated by $\dot=$:
\begin{subequations}\label{Eselrule}\begin{eqnarray}
k+k_i&\dot=&k+k_i+2K\pi/a,\\ m+m_i&\dot=&m+m_i+Mq,\\
\tilde{k}+\tilde{k}_i&\dot=&\tilde{k}+\tilde{k}_i+2\tilde{K}\tilde{\pi},\\
\tilde{m}+\tilde{m}_i&\dot=&\tilde{m}+\tilde{m}_i+\tilde{M}n,
\end{eqnarray}\end{subequations}
where $K$, $M$, $\tilde{K}$ and $\tilde{M}$ are the integers
providing the results in BZ, $(-q/2,q/2]$, $\widetilde{\mathrm{BZ}}$
and $(-n/2,n/2]$, respectively. In the following expressions the
value of the parities $\Pi$ may be $\pm1$ for even and odd states or
0 for all the other states with undefined parity. When this value is
explicitly given (or absent) in expression, the other quantum numbers
are restricted to the compatible values. For given values $(k,m,\Pi)$
and $(k_i,m_i,\Pi_i)$ (or $(\tilde{k},\tilde{m},\Pi)$ and
$(\tilde{k}_i,\tilde{m}_i,\Pi_i)$) the Clebsch-Gordan coefficients
are non-vanishing iff $k_f\dot=k+k_i$ and $m_f\dot=m+m_i+pK$, where
$K=(k+k_i-k_f)a/2\pi$ is an integer, i.e.
$\tilde{k}_f\dot=\tilde{k}+\tilde{k}_i$ and
$\tilde{m}_f\dot=\tilde{m}+\tilde{m}_i$, and $\Pi_f=\Pi\Pi_i$ when
both $\Pi$ and $\Pi_i$ are defined. For $\cZ$ and $\cA$ tubes $p=n$
and $\Pi_f=\Pi\Pi_i$ refers to conservation of each parities
separately. In all these cases the value of the CG coefficient is 1,
\begin{widetext}
\begin{equation}\label{ECGgen}
\braket{k_f,m_f,\Pi_f}{km\Pi;k_im_i\Pi_i}=
\braket{\tilde{k}_f,\tilde{m}m_f,\Pi_f}
{\tilde{k}\tilde{m}\Pi;\tilde{k}_i\tilde{m}_i\Pi_i}=1,
\end{equation}
with the following exceptions:
\begin{enumerate}
\item Chiral tubes, $\tilde{k}\tilde{m}$-numbers:
\begin{equation}\label{ECGCtilde}\begin{tabular}{lll}
&$\braket{\tilde{k}_f,\tilde{m}_f}
{\tilde{k}\tilde{m};\tilde{k}_i\tilde{m}_i-}$&$-1, \mbox{ if
}\tilde{k}<0,\mbox{ or }\tilde{k}=0,\tilde{\pi},\tilde{m}<0;$\\
&$\braket{\tilde{k}_f,\tilde{m}_f}
{\tilde{k}\tilde{m}-;\tilde{k}_i\tilde{m}_i}$&$=-1, \mbox{ if
}\tilde{k}_i<0,\mbox{ or
}\tilde{k}_i=0,\tilde{\pi},\tilde{m_i}<0;$\\
&$\braket{\tilde{k}_f,\tilde{m}_f,\pm}
{\tilde{k},\tilde{m};\tilde{k}_i,\tilde{m}_i}$&$=
\begin{cases}\pm\frac{1}{\sqrt{2}},&\text{$\tilde{k}<0,\mbox{ or }
\tilde{k}=0,\tilde{\pi}, \tilde{m}<0$,}\cr
\frac{1}{\sqrt{2}},&\text{otherwise.}\cr\end{cases}$
\end{tabular}\end{equation}
\item Chiral tubes, $km$-numbers:
\begin{equation}\label{ECGC}\begin{tabular}{lll}
&$\braket{k_f,m_f}{km;k_im_i-}$&$=-1, \mbox{ if }k<0,\mbox{ or
}k=0,m<0,\mbox{ or } k=\pi/a,
m\notin[-\frac{p}{2},\frac{q-p}{2}];$\\
&$\braket{k_f,m_f}{km-;k_im_i}$&$=-1, \mbox{ if }k_i<0,\mbox{ or
}k_i=0,m_i<0,\mbox{ or } k_i=\pi/a,
m_i\notin[-\frac{p}{2},\frac{q-p}{2}];$\\
&$\braket{k_f,m_f,\pm}{k,m;k_i,m_i}$&$=
\begin{cases}\pm\frac{1}{\sqrt{2}},&\text{$k<0$, or $k=0, m<0$ or $k=\pi,
m_i\notin[-\frac{p}{2},\frac{q-p}{2}]$,}\cr
\frac{1}{\sqrt{2}},&\text{otherwise.}\cr\end{cases}$
\end{tabular}\end{equation}

\item Achiral tubes (only the cases with $k=0$ are considered;
$\theta_x$ is the negative step function, being 1 when $x<0$ and
zero otherwise; especially $\theta_{\Pi^s}$ is shorten to
$\theta_s$, for $s=h,v,U$):
\begin{equation}\label{ECGZA}\begin{tabular}{lll}
&$\braket{0,m_f,\Pi^h\Pi^{h_i}}{0,m,B,\Pi^h;0,m_i,\Pi^{h_i}}$&$=-1,
\mbox{ if }m_i<0;$\\
&$\braket{0,m_f,\Pi^h\Pi^{h_i}}{0,m,\Pi^h;0,m_i,B,\Pi^{h_i}}$&$=-1,
\mbox{ if }m<0;$\\
&${\braket{0,m_f,\Pi^v\Pi^{v_i}}{0,m,\Pi^v,-;k_i,m_i,\Pi^{v_i}}}={
\braket{k_i,m_f}{0,m,-;k_i,m_i}} $&$=-1,\mbox{ if }k_i<0;$\\
&$\braket{\pi/a,m_f,-\Pi^h\Pi^{U_i}}{0,m,B,\Pi^h;
        \pi/a,-n/2,\Pi^{U_i}}$&$=-1;$\\
&${\braket{k_f,m_f}{0,m,\Pi^v,\Pi^h;k_i,m_i}}={
\braket{k_f,m_f}{0,m,\Pi^h;k_i,m_i}}$&$=
 (-1)^{\theta_h\theta_{k_i}+\theta_v\theta_{m_i}};$\\
&$\braket{\pi/a,m_f}{0,m,\Pi^h;\pi/a,m_i,\Pi^{U_i}}$&$=
(-1)^{(\theta_h+\theta_{U_i})(\theta_m+\theta_{m_i})};$\\
&$\braket{0,m_f,\Pi^v,\Pi^h\Pi^{h_i}}{0,m,\Pi^h;0,m_i,\Pi^{h_i}}$&$=
\frac{(-1)^{\theta_v\theta_m}}{\sqrt{2}};$\\
&$\braket{\frac{\pi}{a},m_f,\Pi^{U_f}}{0,m,\Pi^h;k_i,0,\Pi^{v_i}}$&$=\frac{
(-1)^{(\theta_{U_f}+\theta_h)\theta_{k_i}+\theta_{v_i}\theta_m}}{\sqrt{2}};$\\
&$\braket{k_f,0,\Pi^{v_f}}{0,m,\Pi^h;\frac{\pi}{a},m_i,\Pi^{U_i}}$&$=\frac{
(-1)^{(\theta_{U_i}+\theta_h)(\theta_{m_i}+\theta_m)+
\theta_{v_f}\theta_m}}{\sqrt{2}};$\\
&$\braket{k_f,m_f,\Pi^{v_f}}{0,m,\Pi^h;k_i,m_i}$&$=\frac{
(-1)^{\theta_h\theta_{k_i}+ \theta_{v_f}\theta_m}}{\sqrt{2}};$\\
&$\braket{k_f,m_f,\Pi^{U_f}}{0,m,\Pi^h;k_i,m_i}$&$=\frac{
(-1)^{(\theta_{U_f}+\theta_h)\theta_{k_i}}}{\sqrt{2}}.$
\end{tabular}\end{equation}
\end{enumerate}
\end{widetext}


\begin{thebibliography}{99}
\bibitem{QT} Note that for SWCT $\tilde{q}=q/n=2\pmod{12}$ is even;
for $\cZ$ and $\cA$ tubes $\tilde{q}=2$. Since $n$ is also the
greatest common divisor of $q$ and $p$, $n$ and $p$ are
simultaneously even only when both $n_1$ and $n_2$ are even.
\bibitem{PI}
The states and representations with $k=\pi/a$ and
$\tilde{k}=\tilde{\pi}/a$ are for short labeled by $\pi$ and
$\tilde{\pi}$, respectively.
\bibitem{HELICAL}
The transition between the two sets of quantum numbers is obtained
calculating the linear and the angular part of the momentum
canonically conjugated to the helix generated by $(C^r_q|na/q)$. The
pair $\tk\tm$ is equivalent to
$k=\tilde{k}-2\pi\tilde{m}r/na+2K\pi/a$, $m=\tilde{m}-Kp+Mq$, while
to the pair $km$ correspond $\tk=k+2\pi
rm/na+2\tilde{K}\tilde{\pi}/a$ and $\tm=m+\tilde{M}n$. The integers
$K$, $M$, $\tilde{K}$ and $\tilde{M}$  provide the quasi momenta from
the proposed intervals. They are independet except that $M$ is a
function of $K$, such that the change of $k$ for $2\pi$ causes the
change of $m$ for $-p$ (for fixed $\tk\tm$). The transition rules
enable also to derive the $\tk\tm$ numbers for the achiral tubes,
which are in the literature treated only with $km$-numbers.
\end{thebibliography}
\end{document}